\documentclass[11pt]{article}
\usepackage{amsmath,amssymb}

\usepackage{epsfig,rotating,pifont}
\usepackage{graphicx}
\usepackage{caption}
\usepackage{subcaption}
\usepackage{dsfont}

\usepackage{hyperref}
\usepackage{cite}

\def\ben{\begin{eqnarray}}
\def\een{\end{eqnarray}}

\newcommand{\bs}{\begin{split}}
\newcommand{\es}{\end{split}}

\newcommand{\be}{\begin{equation}}
\newcommand{\ee}{\end{equation}}
\newcommand{\bea}{\begin{eqnarray}}
\newcommand{\eea}{\end{eqnarray}}
\newcommand{\bg}{\begin{gather}}

\newcommand{\bseq}{\begin{subequations}}
\newcommand{\eseq}{\end{subequations}}

\def\half{\frac{1}{2}}

\def\be{\begin{eqnarray}}
\def\ee{\end{eqnarray}}

\def\p {\partial}
\def\bx {{\bf x}}\def\br {{\bf r}}
\def\bp {{\bf p}}\def\bv {{\bf v}}
\def\bJ {{\bf J}}
\def \bpi{\boldsymbol \pi}

\title{\textbf{The fate of Hamilton's  Hodograph in Special and General Relativity}}

\vspace{2cm}
\author{ \textbf{
 Gary Gibbons$^{1,2,3,4}$ }}

\begin{document}

\maketitle
 
\begin{center}
\emph{$^{1}$ DAMTP, Centre for Mathematical Sciences}\\
\emph{University  of Cambridge , Wilberforce Road, Cambridge CB3 OWA, UK}
\end{center}

\begin{center} 
\emph{ $^{2}$ Laboratoire de Math\'ematiques et Physique Th\'eorique  CNRS-UMR
7350 }\\
\emph{F\'ed\'eration Denis Poisson, Universit\'e Fran\c cois-Rabelais Tours,  }\\
\emph{Parc de Grandmont, 37200 Tours, France} 
\end{center}

\begin{center}
\emph{$^3$ LE STUDIUM, Loire Valley Institute for Advanced Studies,}\\
\emph{ Tours and Orleans, France}
\end{center}

\begin{center}
\emph{$^4$  Department of Physics and Astronomy, }\\
 \emph{University of Pennsylvania, Philadelphia, PA 19104, USA}
\end{center}
{\vspace{-11cm}
\vspace{11cm}}

\begin{abstract}
\noindent {The hodograph of a non-relativistic  particle motion in
Euclidean space is the curve described by its momentum  vector.
For a general
central orbit problem the hodograph is the inverse
of the pedal curve of the orbit, (i.e. its   polar reciprocal), 
rotated through a right angle. 
Hamilton showed that for  the Kepler/Coulomb  problem, the hodograph is
a circle whose centre is in the direction
of a conserved eccentricity vector.  
The addition of an inverse cube law force
induces the eccentricity vector to precess and with it the hodograph. 
The same effect is produced by a cosmic string. If 
one takes the relativistic momentum to define
the hodograph, then  for the  
Sommerfeld (i.e. the special relativistic Kepler/Coulomb problem)  
there is an effective inverse cube force which causes 
the hodograph to precess. 
If one uses Schwarzschild coordinates
one may also define a a hodograph for timelike or null geodesics
moving around a black hole.  Their  pedal equations are  given.
In special cases the hodograph may be found explicitly. 
For example  the orbit of  a photon which starts from the past singularity, grazes the horizon and returns to future singularity  is a cardioid,
its pedal equation is Cayley's sextic the inverse of which is
Tschirhausen's cubic.
It is also shown that  that provided one uses Beltrami coordinates,  
the hodograph for the non-relativistic Kepler problem on hyperbolic space
is  also a circle. An analogous result holds for the the round 3-sphere. 
In an appendix  
the hodograph of a particle freely  moving on a group manifold
equipped  with a left-invariant metric is defined.
}
\end{abstract}

\newpage
\setcounter{tocdepth}{2}
    \tableofcontents
\pagebreak

\newpage

\section{Introduction}

The Kepler or Coulomb  problem is well known to exhibit
many remarkable features and these are often ascribed
to a hidden $SO(3,1)$ $E(3)$ or $SO(4)$   symmetry of phase space
generated by a  conserved, so-called Runge-Lenz, vector.  
Less well known is the relation
to a  fact discovered by Hamilton \cite{Hamilton1,Hamilton2} that the hodograph
of Kepler problem is a circle \cite{Goldstein} and the 
 associated conserved eccentricity vector \cite{Goldstein}.   

It was Hamilton himself who both named and defined the hodograph
associated with the   a non-relativistic motion
of a particle  as the curved described by its
velocity vector $\bv = \frac{d \br}{dt}$ or up to a factor of its mass, its
 momentum vector $\bp= m \bv$ .  
It is not immediately obvious how to generalise this concept 
for a relativistic particle moving in the  flat spacetime of special relativity,
and even less in the  curved spacetime of general relativity.
In the case of special relativity is an obvious guess is to replace
Newtonian time $t$ by propertime $\tau$ along the orbit and we shall 
shortly that this works in a particular case. In general relativity
there is no natural coordinate system in which to define  the velocity
and even if one picks a particular coordinate system, the presence
of curvature will prevent the necessary parallel transport
of the velocity vector required to define a curve as the locus of its
endpoint. Expressed in another way, in a general curved tangent
or co-tangent space, 
there is no natural  projection on to one of its fibres.
\footnote{However, as described in the  appendix, this can be done if the
base space is a group manifold.}  

In the years following  Hamilton's discovery there was a considerable interest
in the hodograph and in particular to the hodographs of central orbit
problems and a number of interesting results were obtained
which in many cases allow  a straightforward  construction of the hodograph
either geometrically, or analytically 
\cite{Original,TaitSteele,Besant,Loney,Ramsey,Routh}.

This suggests that while  searching for a general extension of the
hodograph concept to general relativity, it might be fruitful
to look at those cases, typically spherically symmetric spacetimes,
for which geodesic motion may be reduced to a central orbit problem.
In particular, it seems  worthwhile to look at the 
Schwarzschild solution from this point of view.
There still  remains however some ambiguity as what radial coordinate to use.
It is clear from treatments in standard textbooks that by far the simplest
for our purposes is the traditional Schwarzschild coordinate $r$ and the
 associated radial and tangential velocities 
$\frac{dr}{d \tau}$ and $r^2 \frac{d \phi}{d \tau} $, where $\tau$ 
is propertime for timelike  geodesics  and an affine parameter for
null geodesics.

The organization of the paper is as follows.
In section 1 the definition and  some properties of the 
hodograph in non-relativistic mechanics  are recalled  and 
and generalised  to a relativistic particle moving in flat
spacetime under the influence of Coulomb interaction, a problem
studied by Sommerfeld. It is shown  how relativistic effects    
cause the hodograph to precess. In section 3 the notion of a hodograph 
is extended to  photon orbits in the background of a  Schwarzschild black hole
and its pedal equation given.  For a particular case we find the  
the hodograph curve to be Tschirhausen's cubic. In section 4 
massive particles are treated. In section 5 
It is shown that in Beltrami coordinates, for the Kepler problem
on hyperbolic space the hodograph is also a circle. 
In the  appendix the hodograph of a particle moving freely  on a group manifold
equipped with a left-invariant metric is described in terms of 
generalised Euler equations. Section 6 contains a conclusion
with some future prospects.

 \section{Hodographs for Central Orbits}

We begin by recalling  some  material on central orbits,
and the theory of plane curves, not all of which is as familiar 
today as it was  formerly.
The {\it pedal equation}  of a curve $\gamma$ in the plane, given say in 
polar coordinates $(r,\phi)$ with respect to an origin $S$ by an 
equation  of the form 
$r=r(\phi$),  is  a relation, $p=f(r)$,
between the radial distance $r$ of a point $P$ 
on the curve from the origin and  the perpendicular 
distance $p$ from the origin $S$ 
to the tangent to the curve at the point $(r, \phi)$. Concretely
one eliminates $\phi$ from 
 \ben
p= \frac{r^2 }{\sqrt{r^2 + r^2 (\frac{d r}{d \phi})^2 } } = \label{PPedal}
\frac{1}{ \sqrt{u^2 + (\frac{d u}{d \phi})^2   } }   \,,
\een      
where $u=\frac{1}{r}$.

Now a central orbit with acceleration $F(u)$ towards the centre $S$ 
and angular momentum per unit mass
$h$ satisfies Binet's equation 
\ben
\frac{d^2 u}{d \phi ^2} + u = \frac{F(u) }{h^2 u^2 } 
\een
and has a first integral of the form
\ben
\half \bigl( (\frac{d u}{d \phi})^2   +u^2      \bigr ) 
= \int^u_0 \frac{F(u^\prime) }{h^2 {u ^\prime}^2 } d u^\prime  + C \,   
\een
where $C$ is a constant of integration.
Thus the pedal equation of a central orbit is given by
\ben 
\frac{1}{p^2} = 2 \bigl ( 
\int_0 ^ {\frac{1}{r}}    \frac{F(u)}{h^2 u^2 } du + C \bigr) \label{CentralForce} \,.
\een 
In fact we also have a converse:  if the pedal equation
of a  particle orbit satisfying  Kepler's third law  may be  
cast in the form (\ref{CentralForce})
we may deduce the necessary central force.

\begin{figure}[ht]
\begin{center}
\includegraphics[height=8.5cm]{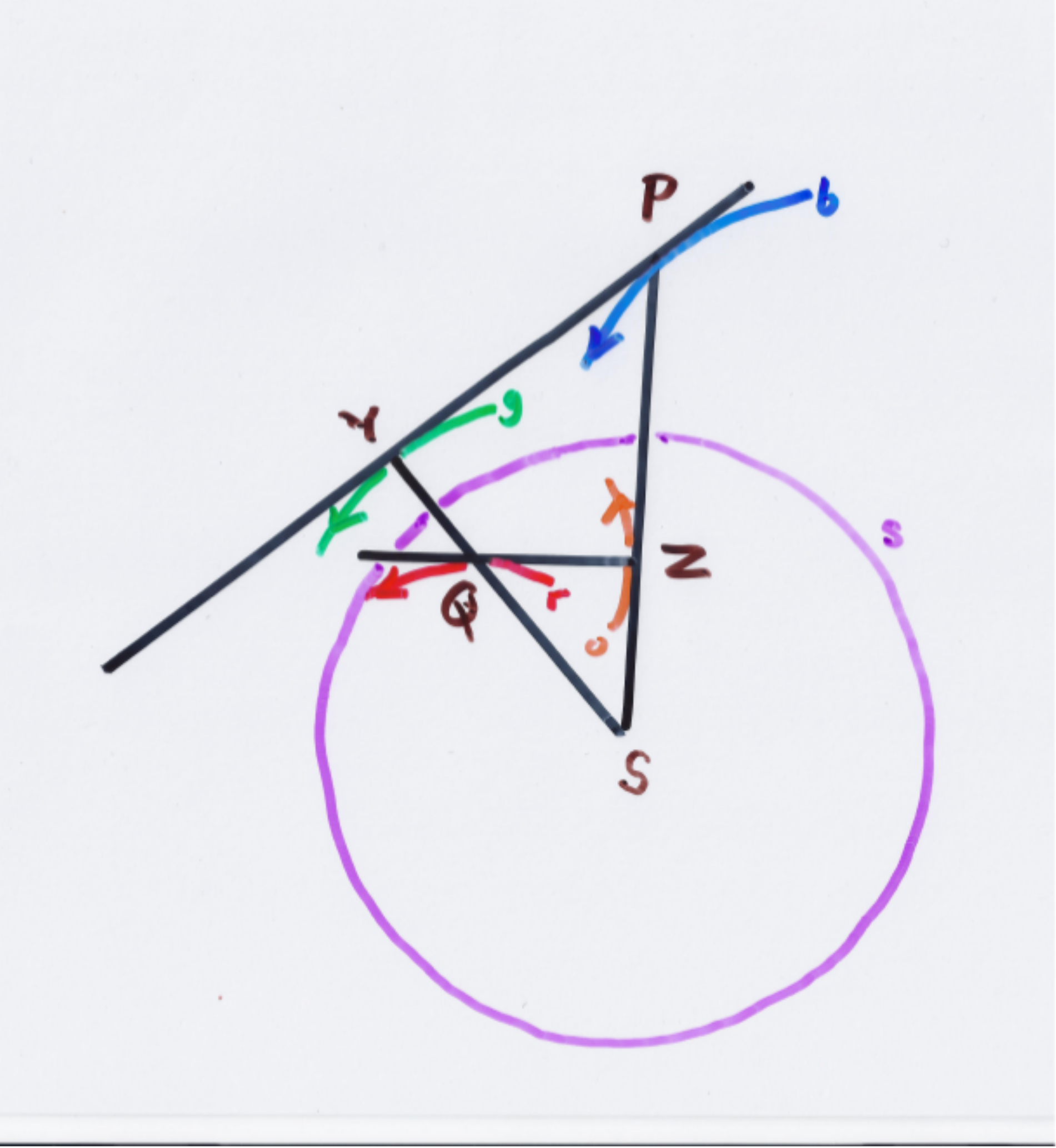}\hspace{0.8cm}
\end{center}
\caption{An orbit  $b$, its pedal curve $g$. 
The inverse of $g$ in the circle $S$ is its    polar reciprocal $r$
which is  the   hodograph
turned through a right angle. The curve  $o$ is the
 pedal curve of the polar reciprocal and its 
inverse in the circle $S$ is the original orbit $b$. }
\label{figdiam}
\end{figure}

The {\it hodograph}  of a particle is the curve swept  out by its velocity 
vector, the velocity vector being parallelly transported to a fixed  origin.
For a central orbit the origin is the centre towards which the force is 
directed. It is then the case \cite{TaitSteele} that {\it the hodograph
is the inverse with respect to the centre 
of the pedal curve of the orbit turned  through a right angle}.
This is illustrated in figure  \ref{figdiam}.

%
%

%

The point  $P$ is the current
point of the orbit $b$ subject to a central force directed
towards $S$, then, since $SY$ is orthogonal to the tangent
$PY$ of the orbit at the point $P$, 
$Y$ is the current point of the  pedal curve $g$.
Now since the force is central we have  $pv=h$, where $p=|SY|$, 
$v$ is the velocity of the orbit at $P$ and $h$ is the
conserved  angular momentum per unit mass. Thus if    
 $Q$ is chosen to lie on $SY$  so that 
$|SY||SQ|= h$,  then $|SQ| =v$. It follows that 
,  since $SQ$ is at right angles to the 
tangent $PY$, the curve $r$ is the hodograph of the orbit $b$ rotated
through a right angle.  More over  $Q$  is  the 
inverse  of  the point $Y$ in the circle  $S$ of radius $\sqrt h$. 

The inverse of the pedal of a curve is also  called 
the {\it reciprocal polar} or just the {\it reciprocal} of the curve
\cite{TaitSteele,Besant,Loney}. The name has its origin in the fact that
the reciprocal polar of the reciprocal polar of a curve is the original
curve (cf \cite{Lamb}).    

This may be seen from the diagram. 
If $Z$ is chosen so that $Z$ is perpendicular to $SP$, than   $Z$ is the 
current point
of the pedal curve $o$ of the the hodograph $r$. 
However,  by construction  the triangles $SZQ$ and $SYP$ are similar so that
\ben
\frac{|SQ|}{|SZ|} = \frac{|SP|}{|SY |} \,, \quad \Rightarrow \quad  
|SP||SZ| = |SQ||SY| = h \,,
\een   
and therefore $P$ is the inverse of the point $Z$ in the circle $s$.
Thus the orbit $b$ is  inverse of the pedal curve $o$ 
of the hodograph and hence
it is the hodograph of the hodograph $r$.

In conclusion  that
{\it The hodograph of a central orbit is the reciprocal
 polar turned through a right angle.}

Thus if $(r_o,p_0),\, (r_p,p_p), \, (r_h,p_h)$ 
are  pedal relations for the orbit, pedal of the orbit, 
and hodograph respectively, we have 
\bea
r_p= p_o\,\qquad p_p = \frac{p_o^2}{r_o} \,, \qquad &\Longleftrightarrow&\qquad
p_o=r_p\,, \qquad r_o= \frac{r_p^2 }{p_p} \nonumber \\
r_h= \frac{h}{p_o}\,\qquad p_h = \frac{h}{r_o} \,, \qquad 
&\Longleftrightarrow&\qquad
p_o=\frac{h}{r_h} \,, \qquad r_o= \frac{h }{p_h} \nonumber \\
r_h= \frac{h}{p_p},\qquad p_h = \frac{h}{p_p} \,,
 \qquad &\Longleftrightarrow&\qquad
r_p= \frac{h}{r_h} \,, \qquad p_p= \frac{h p_h }{r_h^2} 
\,.\label{subst}\eea
  
As an example  the pedal equation of an ellipse with semi-latus rectum
$l$ and  semi-major axis $a$ with respect to its focus
is 
\ben
\frac{l}{p_0^2} = \frac{2}{r_o} - \frac{1}{a} \,. 
\label{ellipse}\een 
The pedal equation of the pedal curve    of the ellipse is 
therefore
\ben
2p_p a =r_p^2 + la \,. \label{Pedal}
\een
Now the pedal equation of a circle of radius $A$ with respect to a point
distance $B$ from the centre of circle is easily seen to be 
\ben
2pA=r^2 +A^2-B^2  \label{circle}
\een  
and therefore the pedal of an ellipse with respect to its centre
is a circle. But the inverse of a circle is a circle
and hence the hodograph is a circle, as claimed by Hamilton.
One may readily check that if one starts
either from (\ref{ellipse}) or (\ref{Pedal}) and uses 
(\ref{subst}) one obtains 
\ben
2 \frac{p_h}{h}= \frac{1}{a} + \frac{lr_h^2}{h^2} \,.\label{hodograph} 
\een 
which is indeed a circle, as found by Hamilton.

More succintly, \cite{Besant} if the pedal equation of the orbit
is $p=f(r)$, that of the hodograph is $\frac{h}{r} = f(\frac{h}{p})$.
and the pedal curve of the orbit has pedal equation $r^2 =p f(p)$.

\subsection{Geometrical and Physical Interpretation of the Hodograph}

Tait and Steele \cite{TaitSteele}  draw attention to some properties
of the hodograph. 
The luminous flux incident on an orbiting planet 
(assuming the inverse square law and ignoring aberration) is proportional
to
\ben
{ 1 \over r^2 } = { 1  \over GMm } |\dot{\bf p}  | =
 {1 \over Gmm} { d  s \over dt}\,.
\een  

{\it Thus the total luminous flux received in a time interval $t$ 
is proportional to the length described by the  the hodograph
in that time.}

Hamilton \cite{Hamilton2} followed by  
Tait  and Steele \cite{TaitSteele}  also  assert
that
\begin{quote} 
It is evident that the path apparently described by a fixed star, in consequence of the 
{\it Aberration} of light, is the Hodograph of the
 Earth's orbit, and is therefore
a circle in a plane parallel to the ecliptic, and of the same dimensions for all stars.
\end{quote}
For  more recent discussions see \cite{Noerdlinger,Shore}
\subsection{ Revolving Orbits,  Global Monopoles and Cosmic Strings}

A useful result in the theory of central orbits, 
due originally to  Newton and  much discussed of late
\cite{DingDong,Ding00,Ding0,Valluri,Ding1}, is that 
{\it Given  a central orbit problem with central acceleration
$F(u)$ and orbits $u=u(\phi, h )$, the orbits of the  associated   central orbit problem with
central acceleration
$F(u) + A u^3 $ are given by $u= u(B \phi, B h) $,
 where $ B=\sqrt{1- \frac{A}{h^2}} $   } 

To see why, recall that  Binet's equation for the original orbit  reads 
\ben
\frac{d^2 u}{ d\phi ^2 } + u = \frac{F(u)}{u^2 h^2}  \,, \label{original}
\een
and for the modified orbit
\ben
\frac{d^2 u}{ d\phi ^2 } + (1- \frac{A}{h^2})  u = \frac{F(u)}{u^2 h^2} \,.  
\een
If $ \tilde \phi = B \phi$ and $\tilde h  =B h$ this becomes
\ben
\frac{d^2 u}{ d {\tilde \phi} ^2 } +  u = 
\frac{ F(u)}
{u^2 {\tilde h}^2} \,,  
\een
which is of the same form as (\ref{original}). 

If successive the apses
of the original orbit  are separated
by an amount $\Delta \phi$ then for the modified orbit they will be separated
by an amount $\Delta (B \phi ) $. In other words the modified orbit
will appear to precess  relative to the original orbit at a rate
of $ (\frac{1}{B}-1)   2 \pi $ per revolution.     

From a geometrical point of view it is as if the original problem on the
equatorial plane with polar coordinates $(r=\frac{1}{u}, \phi) $, with
$0 \le \phi \le 2 \pi$ to a locally flat cone with coordinates 
$(r=\frac{1}{u}, \tilde \phi)$ with  $0 \le \tilde \phi \le 2 \pi B $
with deficit angle $\delta =(1-B) 2 \pi$.
If $A>0$ the deficit angle is a true deficit, $\delta >0$ and
the precession gives rise to  an advance of the apses.     

In general relativity, his situation would  arise
in a an asymptotically flat spherically symmetric static metric 
if it were pierced
by a cosmic string along the an axis of rotational symmetry
and one is considering motion in the orthogonal equatorial plane
(\cite{Hackmann:2009rp}). It would also arise
or for an asymptotically conical
spherically symmetric static metric such as that
of a global monopole \cite{Barriola:1989hx} with spacetime metric
\ben
ds ^2  =- dt ^2 + dr ^2 + B^2  r^2 (d \theta ^2 +  sin ^2 \theta  d \phi ^2 )   
\een

In \cite{Gibbons:1989zv} the {\sl non-relativistic}  Coulomb problem
on such a  cone was investigated. 
The orbits turned out to be precessing ellipses. Moreover
there was a conserved Runge Lenz vector and the hodograph was 
of constant curvature, however because of the deficit angle
neither the orbit nor its hodographs were closed curves
because   of the precession caused by the deficit angle.
In the next section we will see that the  
 {\sl relativistic}  Coulomb problem in {\sl  globally  flat
Minkowski spacetime}  exhibits the same features.

\subsection{Newton's revolving orbit applied 
to the Sommerfeld problem }
The relativistic   Coulomb problem was studied both 
classically (and semi-classically)  by  Sommerfeld
shortly after Bohr's model of the atom in order to explain the fine structure
of the spectral lines of hydrogen. The degeneracy
found in the non-relativistic case is partially  broken
by relativistic effects. 
The complete treatment requires the Dirac equation. Here we deal with
the   the classical motion which, remarkably,  
may be reduced to a central orbit problem with proper time playing the role
of Newtonian time   and solved exactly.  
We start with energy  and momentum conservation 
\ben
\sqrt{m^2c^4 + c^2 \bp^2} -\frac{Ze^2}{4 \pi \epsilon_0 r} = E   
\een
\ben
c^2 \bp ^2 = \bigl( W +\frac{Ze^2}{4 \pi \epsilon_0 r}  \bigr)^2 - m^2 c^4  
\een
\ben
\bx \times \bp = \bJ
\een
where \ben
\bp = m \frac{d \bx}{d \tau}
\een
and $\tau$ is proper time. Thus
\ben
 \bp^2 = m^2 \Bigl( \bigl(\frac{dr}{d \tau}\bigr)^2 + r^2 \bigl
(\frac{d \phi}{d \tau}\bigr) ^2 
  \Bigr) 
\een
\ben
m r^2  \frac{d \phi}{d\tau} = J \,,\qquad u=\frac{1}{r}
\een
\ben
\bigl(\frac{du}{d \phi} \bigr )^2 + u^2  
=   \bigl( \frac{E}{cJ} +\frac{Ze^2}{4 \pi \epsilon_0 cJ }u  \bigr)^2 - \frac{m^2 c^2}{J^2}    \een
where \ben
\alpha ^2 = \frac{e^2}{4 \pi \epsilon_0 c J} \label{fine}   \,.    
\een

Note that semi-classically $J = n \hbar $, $ n \in {\Bbb Z}$. If 
$n=1$  then by (\ref{fine}) $\alpha$ is of course
the usual fine structure constant. Classically we have 
\ben 
\bigl(\frac{du}{d \phi} \bigr )^2 + (1- Z^2 \alpha ^4)
 ( u- \frac{E}{cJ(1 - Z^2 \alpha ^4 ) } ) ^2  
=    \frac{E^2-m^2  c^4 }{c^2J^2} +  \frac{E^2}{c^2J^2 (1 - Z^2 \alpha ^4 ) }
\een
\ben
\tilde \phi = \sqrt{1 - Z^2 \alpha ^4    } \phi  \,
\een
We find
\ben
\bigl(\frac{du}{d \tilde \phi} \bigr )^2 + 
 ( u- \frac{E}{cJ(1 - Z^2 \alpha ^4) ) } ) ^2  
=    \frac{E^2-m^2  c^4 }{c^2J^2 (1 - Z^2 \alpha ^4  )  } + 
 \frac{E^2}{c^2J^2 (1 - Z^2 \alpha ^4 )^2 }
\een

Clearly in the $r,\tilde \phi$ plane  the orbits are precisely  conic sections
with circular hodographs. In the physical $r, \phi$ plane
both the orbit and the hodograph will precess at a steady rate.

\subsection{Non-Separability of Relativistic Euler Problem}

Euler showed that the non-relativistic Kepler  problem with two
fixed centres  is integrable by virtue of an extra constant of the motion.
 In view of the results of \cite{Ding2} who relates the constant  
to Hamilton's eccentricity vector, it seems
worthwhile asking integrability persists in the relativistic case.
In the non-relativistic case, The extra constant of the motion  
is readily found by showing that the Hamilton-Jacobi equation
separates in prolate spheroidal coordinates.  
In fact using prolate spheroidal coordinates it is
easy to see  that the relativistic extension is not separable.

Thus if  $r_1$ and $r_2$ are the distances from two points a distance
$2a$ apart on the axis of symmetry of of a system $u,v,\phi$ 
of prolate spheroidal coordinates the spacetime metric is 
given by 
\ben
ds ^2 = -c^2 dt ^2 + a^2 \bigl( \sinh ^2 u + \sin ^2 v\bigr )\bigl( du^2 + dv ^2\bigr ) + a^2 \sinh ^2 u \sin ^2 v d d \phi ^2 \,,  
\een
and 
\ben
\frac{a}{r_1} = \frac{1}{\cosh u + \cos v} \,,\qquad
 \frac{a}{r_1} = \frac{1}{\cosh u - \cos v} \,.
\een    

The Hamilton-Jacobi equation for a particle of charge
$e$ moving in an electrostatic potential $V$ is 
\bea
&&\frac{1}{a^2} (\p _u S)^2 + (\p_v S) + \frac{1}{a^2}
\bigl( \frac{1}{\sinh^2 u} + \frac{1}{\sin ^2 v} \bigr ) (\p _\phi S)^2 \nonumber \\ 
&=&  \frac{1}{c^2} ( \sinh ^2 u + \sin ^2 v\bigr )   \Bigl
 \{  (E +eV) ^2 - m^2c^4  \Bigr \} \,.  \label{HJ} 
\eea
If 
\ben
V= \frac{q_1}{r_1} + \frac{q_2}{r_s} = 
\frac{1}{a} \frac{(q_1 + q_2 )\cosh u + (q_1 - q_2 )     \cos v}{ \sinh ^2 u + \sin ^2 v} \,,  
\een
the only term on the r.h.s. of (\ref{HJ}) which is not the sum
of a function of $u$ only and a function of $v$ only 
is $\frac{e^2 V^2}{c^2} $ which vanishes in the non-relativistic limit.

Further discussion of the relativistic Euler problem may be found
in \cite{Yoshida,Mirshekari:2010jg} .

\section{Photons 
in the   Schwarzschild metric }  
 
For photons in the Schwarzschild  metric, i.e. for null geodesics
one may reduce the problem to a central orbit problem 
for which  
\ben
\frac{F(u)}{h^2 u^2} = 3M u^2 \,.\label{photon} 
\een
Thus the force is an attraction inversely as
the inverse fourth power of the distance.
It is a striking fact that (\ref{photon}) 
is unaffected by the addition of a cosmological term to the metric
\cite{Islam,Gibbons:2008ru}.

The pedal equation of a photon orbit   is therefore 
\ben
\frac{1}{p^2}  = 2 \bigl( C + \frac{M}{r^3} \bigr )
\een 
Note that for those orbits which reach infinity, we have 
\ben
\frac{1}{p^2} \approx 2C  = \frac{1}{b^2}  
\een
where $b$ is the impact parameter.

A special case is given by $C=0$, in which case 
\ben
\frac{1}{p^2} = 2 M u^3 = \frac{2M}{r^3}  
\een 
which is the pedal equation of the  {\it cardioid}  \cite{Lawrence} p. 118.
\ben
r = M (1+\cos \phi) \,. \label{cardioid}
\een
This has  
parametric equation
\ben
x= M \cos \lambda (1+\cos \lambda) \,,\qquad y= M \sin \lambda (1+ \cos \lambda) 
\een
and Cartesian equation
\ben
(x^2 + y^2 -M x)^2 = M^2 (x^2 + y^2 ) \,.
\een

Thus the photon starts on the past singularity at $\phi=-\pi$ moves outwards
and grazes the horizon at $\phi=0$ and then moves back inwards to the
future singularity at $\phi=\pi$  

The pedal equation of the pedal curve of the photon orbit  is 
in general 
\ben
2Mp^3 = r^4 -2C r^6 \,, 
\een
which, if $C=0$,  is the pedal  equation of {\it Cayley's sextic} \cite{Lockwood} p.  
155. 
\ben
r= 2M \cos ^3 (\frac{\phi}{3})  
\een
whose Cartesian equation is 
\ben
(x^2 + y^2 - 2Mx) ^3 = 27 M^2 (x^2 + y^2 ) ^2 \,.
\een

The inverse of Cayley's sextic is {\it Tschirhausen's cubic} 
whose pedal equation  in units in which $h=1$ is \cite{Lawrence}
\ben
2M r^2 = p^3 \,.
\een
Its  Cartesian equation is
\ben
54M y^2 = (2M-x) (x+ 16M)  \,,
\een
and its parametric equation
\ben
x= 2M ( 1-3 \lambda ^2 ) \,,\qquad y= 2M \lambda  (3-t^2 ) 
\een and   whose polar equation is    
\ben
r=\frac{2M}{ \cos ^3 (\frac{\phi}{3})} \,.\label{Tschirhausen}
\een
Thus {\it Tschirhausen's cubic (\ref{Tschirhausen}) 
turned through a right angle  
is the  hododgraph of the cardioidal photon orbit (\ref{cardioid}) }.

One may continue the chain described above. The pedal equation
of  Tschirhausen's cubic is a parabola with focus at the origin
\ben
\frac{4M}{r}=  1+ cos \phi 
\een 
 and the inverse of this  parabola with respect to the origin
 \ben
\frac{r}{M}=  1+ cos \phi 
\een
is a cardioid. All four curves are {\it sinusoidal spirals} 
of the form $(\frac{r}{b})  ^a = \sin (a\phi)$ 
with $a= \half, \frac{1}{3}, -\frac{1}{3} , - \half $ 
for the cardioid, Cayley sextic, Tschirhausen's cubic, and parabola
respectively.   

In general the hodograph of the photon orbits has pedal equation
\ben
r^2= 2Ch^2 + \frac{2M p^3}{h} \,.
\een

The null geodesics are  given in general  in terms of 
Weierstrass's elliptic function
\ben
\frac{1}{r}= \frac{1}{6M} +  \frac{2}{M}   \frak{p} ( \phi + c) \,,  
\een
where $c$ is a constant of integration.(see e.g. \cite{Gibbons:2011rh})   
and the cardioid is one of three cases where the
Weierstrass function reduces to a  trigonometric or hyperbolic
trigonometric  function. 
The other two have $C= \frac{1}{54 M^2}$  and take the form
\ben
\frac{1}{r}= \frac{1}{3M} - \frac{1}{1+\pm \cosh \phi} \,. 
\een
 
These  start from infinity or the singularity and endlessly encircle 
the circular photon orbit at $r=3M$.

\section{Massive particles  moving
in the   Schwarzschild metric }

 For a massive particle 
\ben
\frac{F(u)}{h^2 u^2} = 3M u^2 + \frac{M}{h^2}\,, \label{tardyon}
\een
and we have a sum of an inverse fourth and inverse square law attraction.
and the pedal equation for both cases is given by  

\ben
r^4-2C r^6   =  2Mp^3 + \frac{2 \epsilon M p r^4}{h^2}  
\label{pedal}  \een
 where $\epsilon =0$ in the massless case and $\epsilon =1$ in the massive case.

Both the massless and massive orbits may be solved in  terms of 
Weierstrass functions \cite{Gibbons:2011rh} and in some cases
are equivalent problems. 
If $v=u+a$ and $\tilde \phi = \sqrt{1-6Ma} \phi $,
and 
\ben
a^2 - \frac{a }{3M} + \frac{1}{3h^2} =0\,, \label{equation}
\een
one finds
\ben
\frac{d^2 v}{ d {\tilde \phi} ^2 } +  v = 
\frac{3  M}{1-6aM} v^2  \,.    
\een 
One may regard (\ref{equation}) either as an equation for $a= a(M,h)$  : 
\ben
a= \frac{1}{6M} \pm \frac{1}{3}  
\sqrt{ \frac{1}{4M^2} -\frac{1}{h^2} } \,
\label{aquation}
\een
or an equation for $h = h(a,M)$ 
\ben
h^2 = \frac{1}{3} \frac{1}{ (1- \frac{1}{6M}) ^2 - \frac{1}{36 M^2 }      } 
\een
In either case {\it given a photon orbit $r=r_p(\phi, M )$, that is a solution
of (\ref{photon}),   
then
\ben
\frac{1}{r} = \frac{1}{r_p(\sqrt{1-6Ma}  \phi, \frac{M}{1-6M a} )   } + 
a
\een
is a solution of (\ref{tardyon})
}.

\section{ Central Orbits in  
Hyperbolic space}
  
The Kepler problem in hyperbolic space has been studied since the
nineteenth  century \cite{Lipschitz,Killing}. More recently   
Higgs \cite{Higgs}
and independently and later   Chernikov \cite{Chernikov}
discussed its remarkable integrability problems.
A recent extensive review is given in \cite{Vozmisheva}, see also
\cite{Gibbons:2006mi} 

The trick is to  use 
Beltrami coordinates ${\bf r}$ , with $r=|{\bf r}|= \tanh \chi$ 
in which the Lobachevsky metric is
\ben
ds ^2 = {d {\bf r} ^2 \over (1-r^2)} + 
{ ({\bf r}.d{\bf r})^2 \over (1-r^2 )^2 }\, 
\een 
and in which  free particles move on straight lines. 
The canonical  momenta are
\ben
{\bf p}= { { \dot {\bf r} } - r^3 {\dot {\hat {\bf r} }} 
 \over (1-r^2 ) ^2 }\,.  
\een
Consider any spherically symmetric potential $V(r)$.
The conserved orbital angular momenta are
\ben
{\bf L}= {\bf r} \times {\bf p} = 
{ {\bf r} \times \dot {\bf r} \over (1-r^2 )}\,.
\een 
The motion lies in a plane and angular momentum conservation
and energy conservation lead to the constancy of 
the angular momentum per unit mass $h$  and the energy $E$ \begin{eqnarray}
h&=& {r^2 \dot \phi  \over (1- r^2) }\,, 
\\
E&=& {1 \over 2} \Bigl [ {\dot r ^2 \over (1-r^2)^2 } 
+{r^2 \dot \phi ^2 \over (1-r^2)  }  
\Bigr ] +V(r)\,.
\end{eqnarray}

Elimination of the time gives
\ben
E= { 1 \over 2} h^2 \Bigl
[ { 1 \over r^4}
 ( {dr \over d \phi} )^2 + { 1 \over r ^2 } \Bigr ]
 + V(r) -{h^2 \over 2}\,.
\een

This is {\it exactly of the same form, for any potential  $V(r)$
as a central  orbit problem  in Euclidean space $ {\Bbb E}^3 $
with flat metric $ds^2 = d \br ^2 $. }.
Indeed if we set $u=\frac{1}{r}$ we have , if $p$ is given by (\ref{PPedal}),
\ben
\frac{1}{p^2} = \frac{1}{h^2} \Bigl( 2E -V(r)   + h^2   \Bigr) 
\een
which is of the same form as (\ref{CentralForce}).

\subsection{The Kepler Problem in Hyperbolic space}

In Lobachevsky space, translations do not commute
but they continue to give conserved quantities if the potential
vanishes. Thus if
\begin{eqnarray}
 {\bf \bpi} &=& {\bf p} -({\bf r}.{\bf p}) {\bf r} =
 {\dot {\bf r} \over (1-r^2)}\,,\\   
{\dot{\bf \bpi}}&=& -\nabla V + ({\bf r} . \nabla V  ) {\bf r} \,. 
\end{eqnarray}

In particular, if we chose for $V$ a spherically symmetric harmonic
function 
\ben
V= \Phi= {q \over 4 \pi r}
\een
we find that
\ben
\dot {\bf \bpi}= { q \over 4 \pi } (1-r^2) { {\bf r} \over r^3}\,
\een
whence we obtain the constant Runge-Lenz vector,
\ben
{\bf K}= {\bf L} \times {\bf \bpi } + {q \over 4 \pi}{\hat {\bf r}}\,,
 \qquad {\dot {\bf K}}=0\,.   
\een

\subsection{The Hodograph is a Circle}

We define this to be the curve swept out by the vector ${\bf \bpi}$.
Since
\ben
{\dot {\bf \bpi}} = { q \over 4 \pi} { {\bf r} \over r^3} (1-r^2) \,,  
\een
so that the  the tangent vector of the hodograph is in the radial direction
and the angle $\tilde \psi$ 
the tangent makes with a fixed direction is $\phi$. 
Moreover if $\tilde s$ is
the arc-length along the hodograph    \ben
{ d \tilde s \over dt} = |\dot \pi| =  { q \over 4 \pi r^2} (1-r^2)\,.
\een

Now the radius of  curvature $ \tilde \rho $  of the hodograph is given by
\ben
\tilde \rho ={d \tilde s \over  d \tilde \psi} ={d \tilde s  \over   d  \phi}
= { d \tilde s \over dt} { d t \over d \phi} = { q \over 4 \pi h }\,. \een
  
{\it Thus the hodograph is a plane curve with a constant 
radius of curvature, i.e. a circle.}

\section{Conclusion}

In this paper, The extension of   of Hamilton's 
notion of a hodograph to cover a particle moving
in a curved background, possibly relativistically,
has been studied. In flat space time
the extension to include relativistic effects
appears to present no great problems, even though
relativistic effects may lead quantities which
are conserved non-relativistically no longer being conserved.    
If in a curved spacetime  the problem reduces, 
on choosing suitable coordinates,
to a central orbit problem one may still define 
 the hodograph  in  straightforward way.
The case of geodesics in the  Schwarzschild aolution has  
been treated in detail but the procedure adopted would work
for any spherically symmetric  static metric.
Of course in that case, there is some freedom
in the choice of radial coordinate and the example
of hyperbolic space shows that an appropriate choice
can lead to dramatic simplifications.  

Les obvious is how to proceed if the metric
is not spherically symmetric. In the case of 
free motion on a group manifold one 
may regard the analogue of Euler equations for a top
as giving the hodograph. The case of hyperbolic
space, which is a coset rather than a group
manfold suggests a possible route to explore
in the future. Another question for future study
would be the Sommerfeld problem on hyperbolic space.

\section{Acknowledgement}

I would like to thank Peter Horvathy for his interest in this work
and also  Thanu Padmanabhan who suggested to me 
some years ago 
that the hodograph for Coulomb motion  on hyperbolic space 
might  be a circle.

\section{Appendix: The Hodograph on a  Group Manifold}

We start by giving  general treatment of Hamiltonian
mechanics on a group manifold, obtaining the Euler equations
and the equations for the time dependence of the coordinates
on the group manifold.

Given a Lie group $G$, coordinates $x^\mu$, i.e. group elements
$G \ni g=g(x^\mu)$, and  left and right invariant Cartan-Maurer forms
\ben
 g^{-1} dg= \lambda ^a {\bf e}_a  \,,\qquad  dg g^{-1} = 
\rho ^a {\bf e}_a 
\een
with ${\bf e}_a$ a basis for the Lie algebra $\frak{g}$
such that  
\ben
[{\bf e}_a ,{\bf e}_b ] =C_a\,^c\,_b  \,{\bf e}_c
\een
\ben
d \lambda ^c = -\half C_a\,^c\,_b \,\lambda ^a \wedge \lambda ^b 
\,,\qquad d \rho ^c = \half C_a\,^c\,_b \,  \rho ^a \wedge \rho ^b
\een we pass to the co-tangents space $T^\star G= G \times
\frak{g} ^\star$ with Darboux  coordinates $( x^\mu, p_\mu ) $.    

The left and right  invariant 
vector field $L^\mu_a$ and $R^\mu _a $ dual to
$\lambda^a _\mu \,,  \rho ^a _\mu $  repectively, 
\ben
\lambda^a _\mu L _b^\mu =
 \delta ^a_b \,,\qquad 
\rho^a _\mu R _b^\mu = \delta ^a_b \,,
\een
satisfy  
\ben
[L_a, L_b ] = C_a\,^c\,_b \, L_c  
\qquad [R_a, L_b ]=0\,, \qquad [R_a, R_b ] = -C_a\,^c\,_b \,  R_c 
\een
and generate  right  and left  translations on $G$.
Quantum mechanically one often inserts $i$'s so that
if $\hat R_a ={1 \over i} R_a$, $\hat L_a ={1 \over i} L_a   $  then
\ben
[ \hat R_a, \hat R_b ] = i C_a\,^c\,_b \, \hat R_c\,, \een
\ben
 [ {\hat L} _a, {\hat L}_b  ]   = -i C_a\,^c\,_b \,{\hat L} _c \,.
\een

We may define moment maps into $\frak{g}^\star $, the dual
of the Lie algebra,
\ben
M_a= p_\mu L^\mu _a\,,\qquad N_a= p_\mu R^\mu _a\,,
\een
with Poisson brackets
\ben
\{M_a, M_b \} = -C_a\,^b\,_c M_b  
\qquad \{M_a, N_b\}=0\,, \qquad \{N_a, N_b \} = C_a\,^b\,_c M_b 
\een 
which generate the lifts of right and left translation
to $T^\star G$.  
A Hamiltonian $H=H(x^\mu, p_\mu)$  which is left-invariant
satisfies
\ben
\dot N_a = \{N_a , H \}=0\, ,
\een
and so the moment maps $N_a$ are constants of the motion.
By contrast, the moment maps generating right actions,  $M_a$,
are time-dependent 

\ben
\dot M_a = \{M_a , H \} \ne 0\, ,
\label{Euler} \een

A left-invariant Lagrangian may be constructed from
combinations of left-invariant velocities or angular velocities
\ben
\omega^a =\lambda ^a _\mu  \dot  x ^\mu  
\een
Thus the Hamiltonian is a combination of
the momenta maps $M_a$,
\ben
H= H(M_a) \,
\een
Thus (\ref{Euler}) provide an autonomous 1st order  system
of ODE's on $\frak{g}^\star$ for the moment maps 
$M_a$ called the {\it Euler equations}. 
To obtain the motion on the group, one uses the 
equation
\ben
\dot x ^\mu = {\p H \over \p p_\mu}\, 
\een
Now 
\ben
p_\mu = M_a \lambda ^a _\mu 
\een
and so
\ben
\dot x^\mu = L_a ^\mu {\p H \over \p M_a} \,.
\een 
 
The method described above can reasonably be called {\it hodographic}.
Hamilton \cite{Hamilton1,Hamilton2} defined the {\it hodograph} 
of a particle  motion ${\bf x}= {\bf x}(t)$ in   ${\Bbb E} ^3$ as the  
the curve  described by 
velocity vector ${\bf v}(t)={d {\bf x} \over dt } $,
a construction very similar to the Gauss map for surfaces in ${ \Bbb E} ^3$.
Hamilton then discovered \cite{Hamilton1,Hamilton2} the elegant result that the hodograph for 
Keplerian motion
is a circle.

Since velocity space and momentum space are naturally identified in this
case  we may think about the motion in phase space $T^\star {\Bbb E} ^3
= ( {\bf x}, {\bf p} )$ , 
and then 
observed that because ${\Bbb E} ^3$ is flat, there is, in addition to the
standard {\it vertical}  projection 
$ ( {\bf x}, {\bf p} ) \rightarrow ( {\bf x}, 0 )$, 
 a well defined  {\it horizontal} map  or 
{\it hodographic}   projection $ ( {\bf x}, {\bf p} ) \rightarrow
( 0, {\bf p}) $. For a general configuration space
$Q$, the co-tangent manifold $T^\star Q$, will not admit
a well-defined horizontal projection. However  if $Q=G$, a group
manifold, then it does, and the Euler  equations 
govern the motion of the hodograph.

\end{document}